\def\aa{{A\&A}}

\def\apj{{ApJ}}
\def\apjs{{ApJS}}

\def\mnras{{MNRAS}}

\documentclass[cup5b]{caps}
\usepackage{psfig}
\usepackage{graphicx}
\usepackage{amssymb}
\usepackage{ociwsymp3e}  

\HeadText{D.D. Kocevski et al.}  

\def\plottwo#1#2{\centering \leavevmode
\includegraphics[width=.49\columnwidth]{#1} \hfil
\includegraphics[width=.49\columnwidth]{#2}}

\begin{document}

\pagenumbering{arabic}
\renewcommand{\thefootnote}{\arabic{footnote}}

\author[]{D.D. Kocevski$^{1}$, H. Ebeling$^{1}$, C.R. Mullis$^{2}$\\
\\
(1) IfA, University of Hawaii, Honolulu, HI, USA \\
(2) European Southern Observatory, Garching, Germany } 
%
%

\chapter{Clusters in the Zone of Avoidance}

\begin{abstract}
The Clusters in the Zone of Avoidance (CIZA) project is producing the first statistically complete catalogue of X-ray selected galaxy clusters behind the Galactic plane.  Current optically selected cluster catalogues have excluded a wide band of sky centered on the Milky Way ($| b |\leq 20^{\circ}$) due to the high extinction and stellar obscuration at low Galactic latitudes.  An X-ray search for cluster candidates overcomes the problems faced by optically selected catalogues, hence providing the first opportunity to explore the sky in the traditional Zone of Avoidance; principally, the Great Attractor region.  When added to other X-ray selected cluster surveys such as the BCS and REFLEX, CIZA will complete the first unbiased, all-sky cluster catalogue.  We here report on the progress of the CIZA survey as well as preliminary results of a dipole analysis which utilizes the current CIZA sample.
\end{abstract}

\section{Introduction}
Clusters of galaxies are the most massive gravitationally collapsed structures known to exist and are significant to the dynamics of the local universe since they act as accelerators for large-scale flows.  Traditional optical cluster searches (Abell et al. 1989) suffer from severe extinction and stellar obscuration in the direction of the Milky Way (MW), leading to catalogues with poor coverage in a $40^{\circ}$ wide strip centered on the plane of the Galaxy, known as the Zone of Avoidance (ZOA).  This is particularly troubling since Tammann \& Sandage (1985) suggested that large-scale structures obscured by the ZOA could have a significant effect on the peculiar motion of the Local Group (LG).  More recently, several studies have found renewed evidence for a significant bulk motion toward a vertex behind the MW (Riess et al 1997; Hudson et al. 1999), rekindling the idea of a Great Attractor (Lynden-Bell et al. 1988).  A cluster or set of clusters massive enough to produce such a flow has yet to be conclusively identified, mainly because of the attractor's location behind the MW.  A variety of techniques have been used to reconstruct the ZOA, ranging from a uniform filling (Strauss \& Davis 1988; Lahav 1987) to a spherical harmonic approach which extends structures above and below the plane into the ZOA (Tini-Brunozzi 1995).  The value of these reconstruction techniques is limited if the MW does indeed obscure dynamically significant regions, as has been suggested.


An X-ray search for clusters overcomes many of the problems posed by optical searches. The advantage presented by X-ray observations is three-fold:  first the level of extinction is based only on the X-ray absorbing equivalent Hydrogen column density, $n_{H}$, and tends to be moderate at all Galactic latitudes compared to the severe extinction in the optical.  The distribution of $n_{H}$ in the Galactic plane is shown in Figure 1; note the large regions of relatively low absorption ($n_{H} < 5\times 10^{21}$cm$^{-2}$) well inside the traditional ZOA.  Secondly, the number of objects emitting at X-ray wavelengths in the MW is low, making X-ray searches immune to source confusion and obscuration problems that plague optical searches.  Lastly, X-ray emission from clusters is more peaked at the gravitational center of a cluster than the projected galaxy distribution, effectively eliminating projection effects that arise in optical surveys.  The most extensive cluster catalogues produced via an X-ray search thus far are the Brightest Cluster Sample (BCS, Ebeling et al. 1998, Ebeling et al. 2000) and the ROSAT-ESO flux limited X-ray sample (REFLEX, B\"{o}hringer et al. 2001), which together cover most of the extragalactic sky down to an X-ray flux of roughly $3\times 10^{-12}$ erg cm$^{-2}$ s$^{-1}$ in the 0.1-2.4 keV band.  Although comprehensive, both the BCS and REFLEX surveys avoided the MW and are therefore limited to Galactic latitudes greater than $|b| \sim 20^{\circ}$.  Due to this incompleteness, there existed no extensive catalogue of clusters in the ZOA prior to the start of the CIZA project.

\section{The CIZA Project}
The CIZA survey (named for Clusters in the Zone of Avoidance) has begun to lift the veil cast by the MW by performing the first systematic X-ray selected search for clusters at low Galactic latitude.  Our strategy is to use the ROSAT Bright Source Catalogue (BSC, Voges et al. 1999) to select cluster candidates and follow these targets up with optical observations.  We apply three selection criteria to the BSC to obtain our initial target list: 1.) sources must lie near the plane of the Galaxy, $| b | < 20^{\circ}$, 2.) they must have an X-ray flux$^{1}$\footnotetext[1]{All reported CIZA fluxes are in the 0.1-2.4 keV band} greater than $1\times 10^{-12}$ erg cm$^{-2}$ s$^{-1}$ and 3.) their spectral X-ray hardness ratio must exceed a preset threshold value$^{2}$\footnotetext[2]{The minimum hardness ratio cut depends on location in the plane; see Ebeling, Mullis \& Tully (2002) for details} to discriminate against softer, non-cluster sources.  The resulting list is cross-correlated with the NED, SIMBAD and DSS databases to remove any known clusters and obvious non-clusters (i.e. stars, supernova remnants, AGN/QSO, etc).  The remaining sources are then subjected to a comprehensive imaging survey in the R band.  If a source is confirmed to be a cluster, it is then targeted for spectroscopic observations to measure the redshift of at least two cluster members.  Clusters discovered in dynamically interesting regions such as the GA are further followed up with observations by the XMM X-ray observatory.



Since 1998 an observational campaign has been underway to construct the CIZA catalogue.  The survey is split into four X-ray flux limited sections, with the brightest and faintest subsamples having a flux limit of $5\times 10^{-12}$ and $1\times 10^{-12}$ erg s$^{-1}$ cm$^{-2}$, respectively (see Table 1).  Using the UH 2.2m, CTIO 1.5m and NTT 3.5m telescopes for R band imaging and spectroscopic follow-up, CIZA has thus far discovered 150 clusters, 124 of which come from the two brightest subsamples that are effectively complete. The distribution of these clusters on the plane of the Galaxy is shown in Figure 1.  Arguably the most significant finding of the CIZA survey thus far has been the discovery of several, previously unknown galaxy clusters in the Great Attractor region.  Located at similar or greater redshifts than A3627, these systems indicate that the Great Attractor may be a more extended structure than previously thought (Ebeling, Mullis \& Tully 2002).  We have been awarded time on the XMM telescope to image three clusters in and around the Great Attractor region and we anticipate measuring their masses and hence evaluating their dynamical significance in the near future.  With additional cluster candidates awaiting spectroscopic confirmation, CIZA has only begun to shed light on the sky behind the MW.

\begin{figure}
   \centering
   \includegraphics[width=11.7cm,angle=0]{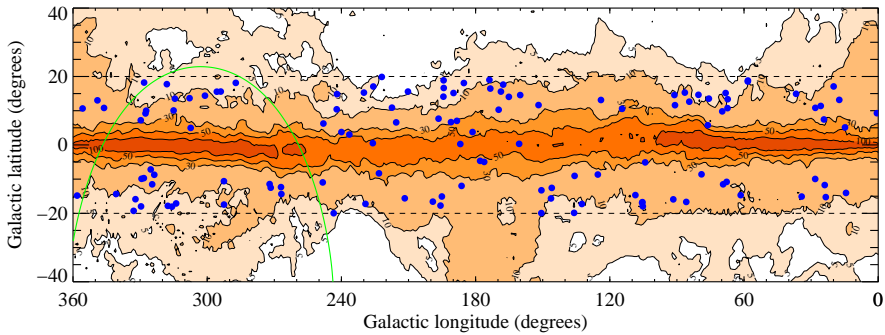}
   \caption{The locations of the 124 CIZA clusters with BSC fluxes greater than $3\times 10^{-12}$erg cm$^{-2}$ s$^{-1}$ overlaid on contours of Galactic $n_{H}$ column density (in units of $10^{20}$ cm$^{-2}$).  The optical ZOA is marked by the dashed lines.  The green line denotes the part of the sky not easily observable from Mauna Kea, hence requiring observations in the southern hemisphere.}
\end{figure}

Predictions for the statistical completeness of each CIZA subsample can be calculated by folding the BCS X-ray luminosity function (Ebeling et al. 1997) through the CIZA selection function (Ebeling, Mullis \& Tully 2002).  This effectively allows us to calculate the number of clusters expected from the BSC catalogue at a certain redshift. These predictions are shown in Figure 2 along with the current CIZA sample in each flux limit.  Observations thus far have concentrated on the X-ray brightest targets; this fact is reflected in the completeness of the two brightest subsamples, B1 \& B2. These subsamples are 70.2\% \& 68.6\% complete out to z$\sim$0.3 and 100\% complete out to z$\sim$0.075.  The completeness of each subsample is given in Table 1.  The remaining incompleteness in B1 and B2 come from the difficulty to spectroscopically confirm targets at higher redshifts.  We also expect to miss a few nearby systems at very low Galactic latitudes where extreme stellar densities prohibit the optical confirmation of cluster candidates.  Hence we do not anticipate increasing the completeness of these subsamples and use this as an upper estimate of the completeness expected in our two fainter subsamples. 

\begin{figure}
   \centering
   \plottwo{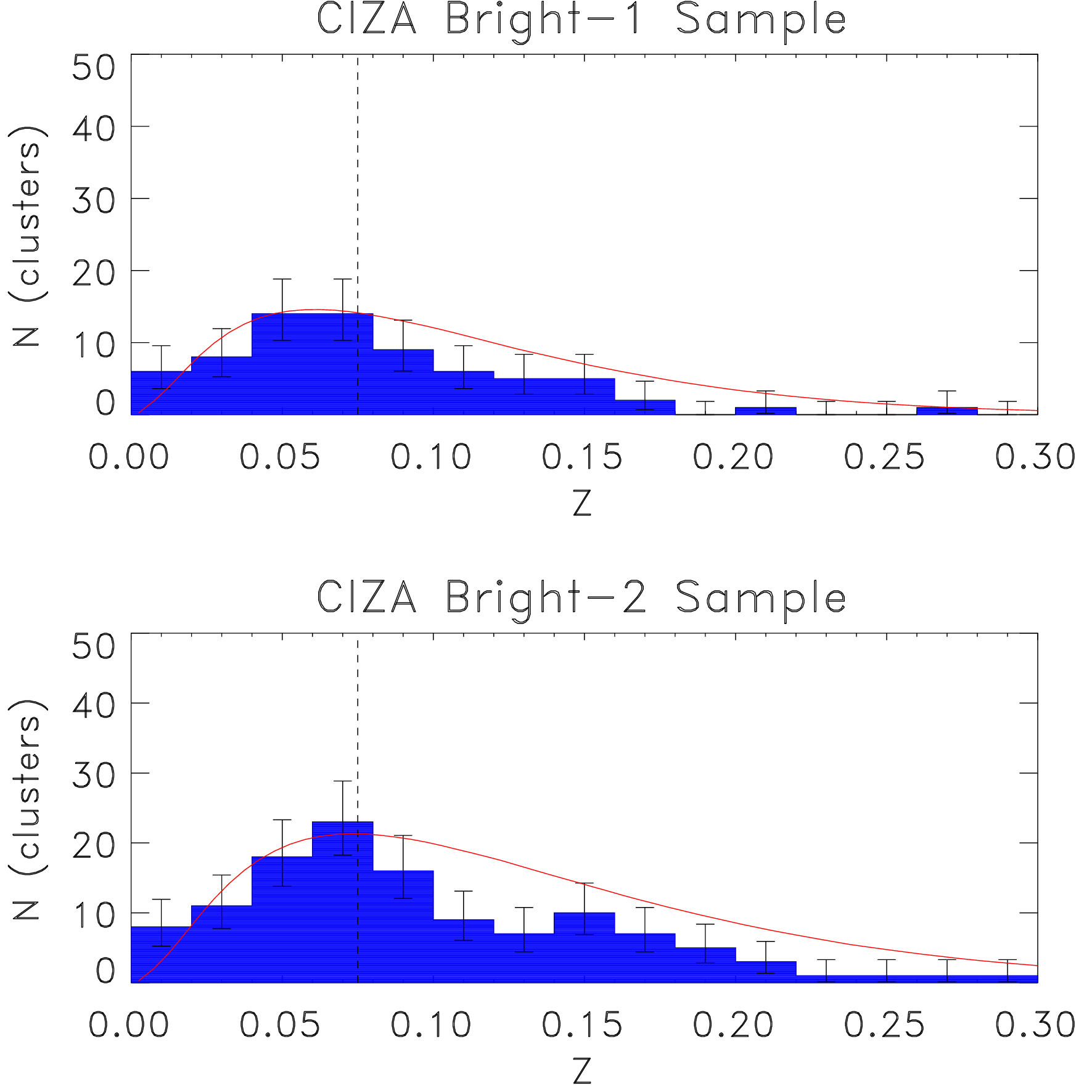}{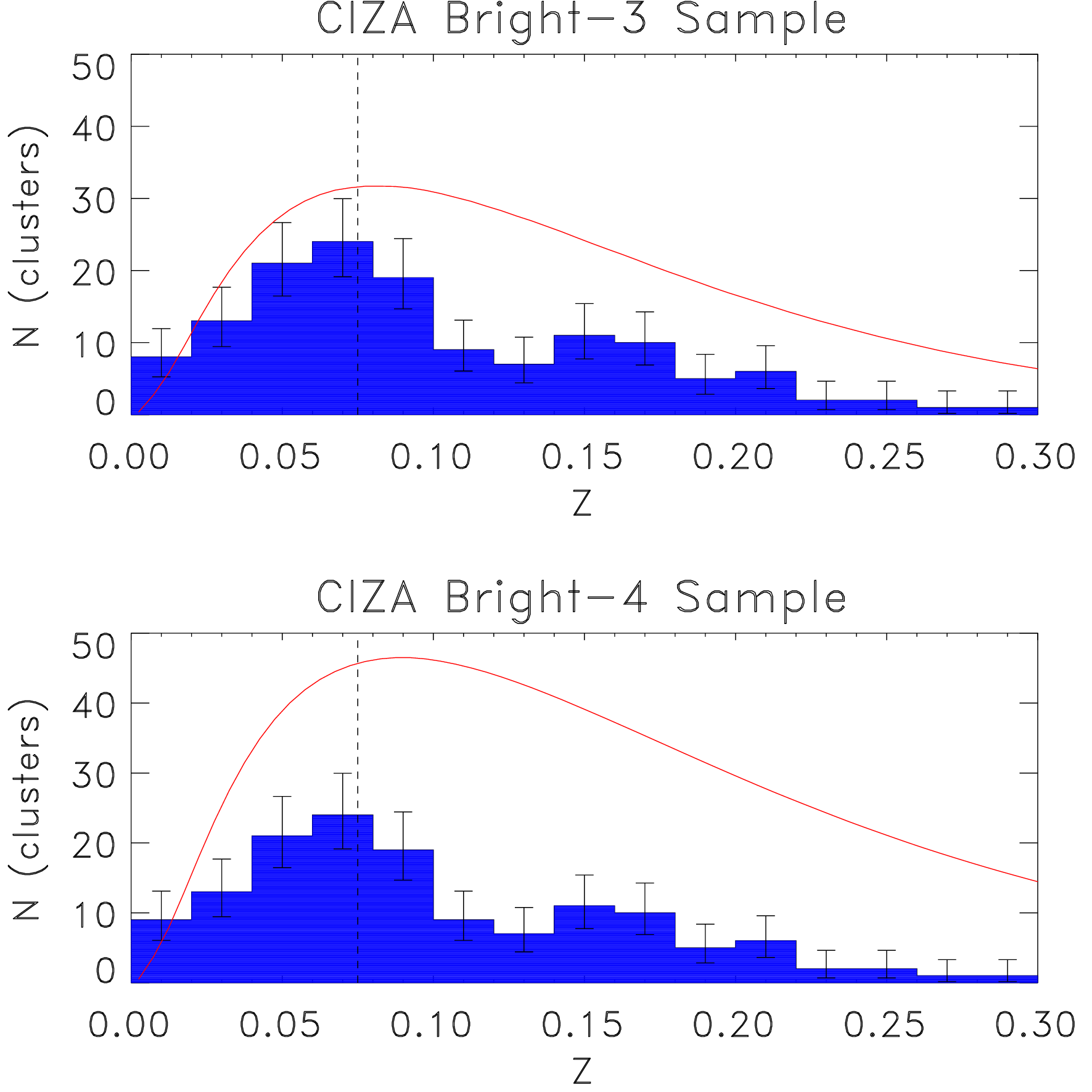}	
   \caption{Completeness levels for the four CIZA subsamples. The blue histogram shows clusters discovered and the red curve represents the number of clusters predicted from the BSC luminosity function and the CIZA selection function.  The flux limits for the B1,B2,B3 and B4 samples are 5,3,2 and $1\times10^{-12}$ erg s$^{-1}$ cm$^{-2}$, respectively.}
\end{figure}

\begin{table}[h]
 \caption{Current completeness estimates for subsamples of the CIZA survey.}
   \begin{tabular}{c|ccccc}
    \hline \hline
          & Flux Limit  &  BSC  &  Expected  &  Discovered  &  Completeness \\
Subsample & (erg s$^{-1}$ cm$^{-2}$) & Targets & Clusters$^{\dag}$ &  Clusters$^{\dag}$&   (\%)$^{\dag}$\\ \hline
B1 & 5$\times 10^{-12}$ & 481 & 101 (39) & 71 (41) & 70.3 (100)\\ 
B2 & 3$\times 10^{-12}$ & 336 & 79 (13) & 53 (17) & 67.1 (100)\\
B3 & 2$\times 10^{-12}$ & 387 & 132 (22) & 18 (6) & 13.6 (27.2)\\
B4 & 1$\times 10^{-12}$ & 697 & 220 (29) & 1 (1) & 0.4 (3.4)\\
    \hline \hline
    \multicolumn{6}{l}{\footnotesize{$^{\dag}$ Values in parentheses are for clusters with z $<$ 0.075.}}
   \end{tabular}
\end{table}

\section{Dipole Analysis}

With the compilation of the CIZA catalogue, it will be possible to study the dynamical effects that nearby clusters have on the LG without the need for reconstruction methods to fill the ZOA.  According to the linear theory of gravitational instability, the peculiar velocity of the LG can be related to the gravitational acceleration induced by the mass distribution surrounding it.  In the linear biasing paradigm (Kaiser 1984), this relationship can be written as

\begin{equation}
\textit{\textbf{v}}_{p} \hspace{.1in} = \hspace{.1in} \frac{H_{o}\Omega^{0.6}_{0}}{4\pi \bar{n} b} \int \frac{n(r)}{r^{2}} \textit{\textbf{\^{r}}} \hspace{.1in} = \hspace{.1in} \beta\textit{\textbf{D}} \
\end{equation}
where $\beta = \Omega^{0.6}_{0}/b$ and \emph{b}, the biasing parameter, is the linear factor that relates the surrounding mass distribution to the tracers that sample it.  Traditionally, the gravitational acceleration induced by these tracers is estimated by measuring the dipole moment of their distribution, while compensating for known systematic effects in the sample.  This dipole analysis has been extensively applied to the LG, since it has a peculiar velocity that is well determined from the dipole structure of the Cosmic Microwave Background (CMB) (Kogut 1993).  Given $\textit{\textbf{v}}_{p}$ and measuring $\textit{\textbf{D}}_{cl}$ from a cluster sample, we can obtain an estimate of the cosmological $\beta$ parameter as well as measure the depth at which anisotropies in the cluster distribution cease to effect LG dynamics (the convergence depth, $R_{conv}$).

For the purpose of this analysis we combine CIZA clusters that have an X-ray flux greater than $5\times10^{-12}$ erg s$^{-1}$ cm$^{-2}$ to the X-ray Brightest Abell-type Cluster catalogue (XBACs, Ebeling et al. 1996) to create an all-sky sample of X-ray clusters.  To remove any systematics that would affect the measured dipole, we apply statistical weights to each cluster in the combined sample.  The primary weight is the inverse of the sample's selection function, which compensates for the non-detection of intrinsically less luminous members with increasing distance.  An additional weight is applied to XBACs clusters to correct for the sample's known deficiency of low luminosity, low redshift systems (Ebeling et al. 1998).  Finally all clusters are given a weight based on their relative mass, as estimated by $M \propto L^{3/4}_{x}$ (Allen et al. 2003).

\begin{figure}
   \centering
   \includegraphics[width=11.7cm,angle=0]{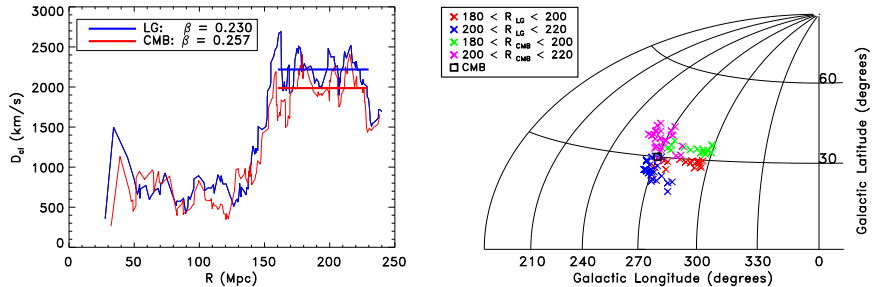}
   \caption{The X-ray cluster dipole amplitude versus distance (\emph{left}) and the dipole direction in galactic coordinates (\emph{right}).  The amplitude is derived in both the LG and CMB frames, since calculations in each frame overestimate and underestimate, respectively, the true cluster dipole amplitude.   Note the overall agreement between the cluster dipole direction and the location of the CMB dipole; at 200 $h^{-1}$ Mpc the dipole directions differ by only $2.5^{\circ}$}
   \label{My Label}
\end{figure}

Preliminary results of our dipole analysis are shown in Figure 1.3.  We find that the cluster distribution becomes isotropic with respect to the LG at a distance of 160 $h^{-1}$ Mpc, as is indicated by the flattening of the dipole amplitude.  This is in agreement with the results of previous analyses that utilized XBACs without the CIZA supplement (Plionis \& Kolokotronis 1998), but differ from results obtained using a variety of galaxy catalogues (Strauss et al. 1992, Hudson 1993), indicating that galaxy samples alone do not probe the mass fluctuation field deep enough to account for all the anisotropies that effect the LG's dynamics.  We also find excellent agreement in the dipole direction as measured by both the X-ray cluster distribution and the CMB anisotropy (Lynden-Bell \& Lahav 1988); at 200 $h^{-1}$ Mpc the dipole directions differ by only $2.5^{\circ}$.  Furthermore, averaging the dipole amplitude obtained in the LG and CMB reference frames (see Tini-Brunozzi et al. 1995) and comparing it to the LG peculiar velocity inferred from the CMB dipole, we find a value of $0.243\pm 0.002$ for the cosmological $\beta$ parameter.  These results and a more detailed analysis of the CIZA contribution to the LG's dynamics can be found in Kocevski, Mullis \& Ebeling (2003).


\begin{thereferences}{}

\bibitem{}
Abell, G.~O., Corwin, H.~G., \& Olowin, R.~P. 1989, \apjs, 70, 1
\bibitem{}
Allen et al. 2003, arXiv:astro-ph/0208394
\bibitem{}
B\"{o}hringer, H., et al. 2001, \aa, 369,826
\bibitem{}
Ebeling, H.,Voges, W.,B\"{o}hringer, H.,Edge, A.~C.,Huchra, J.~P. \& Briel, U.~G. 1996, \mnras, 281, 799
\bibitem{}
Ebeling, H., Edge, A.~C., Fabian, A.~C., Allen, S.~W., Crawford, C.~S., B\"{o}hringer, H. \& Huchra, J.~P. 1997, \apj, 479, L101
\bibitem{}
Ebeling, H., Edge, A.~C., B\"{o}hringer, H., Allen, S.~W., Crawford, C.~S., Fabian, A.~C., Voges, W. \& Huchra, J.~P. 1998, \mnras, 301, 881 
\bibitem{}
Ebeling, H., Edge, A.~C., Allen, S.~W., Crawford, C.~S., Fabian, A.~C. \& Huchra, J.~P. 2000, \mnras, 318, 333
\bibitem{}
Ebeling, H., Mullis, C.~R. \& Tully, R.~B. 2002, \apj, 580, 774
\bibitem{}
Hudson, M. 1993, \mnras, 265, 72
\bibitem{}
Hudson, M.~J., Smith, R.~J., Lucey, J.~R., Schlegel, D.~J., Davies, R.~L. 1999, 512, L79  
\bibitem{}
Kaiser, N. 1984, \apj, 284, L9
\bibitem{}
Kocevski, D.~D. , Mullis, C.~R. \& Ebeling H. 2003, \emph{in prep}
\bibitem{}
Kogut, A. et al. 1993, \apj, 419, 1
\bibitem{}
Lahav O. 1987, \mnras, 225, 213
\bibitem{}
Lynden-Bell, D., et al. 1988, \apj, 326, 19
\bibitem{}
Lynden-Bell, D. \& Lahav, O. 1988, in Large Scale Motions in the Universe: A Vatican Study Week, ed Rubin V.~C., Coyne G. (Princeton: Princeton Univ. Press) 
\bibitem{}
Plionis, M. \& Kolokotronis, V. 1998, \apj, 500, 1
\bibitem{}
Riess, A.~G., Davis, M., Baker, J. \& Kirshner, R.~P. 1997, \apj, 488, L1
\bibitem{}
Strauss, M.~A., Yahil, A., Davis, M., Huchra, J.~P. \& Fisher, K. 1992, \apj, 397, 395 
\bibitem{}
Strauss, M.~A. \& Davis, M. 1988, in Large Scale Motions in the Universe: A Vatican Study Week, ed Rubin V.~C., Coyne G. (Princeton: Princeton Univ. Press) 
\bibitem{}
Tini-Brunozzi, P., Borgani, S., Plionis, M., Moscardini, L. \& Coles, P. 1995, \mnras, 277,1210
\bibitem{}
Tammann, G., \& Sandage, A. 1985, \apj, 294, 81
\bibitem{}
Voges, W., et al. 1999, \aa, 349, 389

\end{thereferences}
\end{document}